\documentclass{aa}
\usepackage{epsfig}
\def\ltsima{$\; \buildrel < \over \sim \;$}
\def\simlt{\lower.5ex\hbox{\ltsima}}
\def\gtsima{$\; \buildrel > \over \sim \;$}
\def\simgt{\lower.5ex\hbox{\gtsima}}
\def\src{IRAS12393+3520}

\begin{document}
\thesaurus{03(13.25.2; 				
		11.01.2; 			
		11.19.1; 			
		11.09.1 IRAS12393+3520; 	
		11.09.4; 			
		11.14.1)}			

\title{A direct view of the AGN powering \src }
\author{M.Guainazzi\inst{1,2},  M.Dennefeld\inst{3}\thanks{Visiting
astronomer, Observatoire de Haute-Provence (OHP), CNRS, France}, 
L.Piro\inst{4}, T.Boller\inst{5}, P.Rafanelli\inst{6}, M.Yamauchi\inst{7}}

\institute{
{Astrophysics Division, Space Science Department of ESA, ESTEC, Postbus 299,
NL-2200 AG Noordwijk, The Netherlands}
\and
{XMM SOC, VILSPA, ESA, Apartado 50727, E-28080 Madrid, Spain}
\and
{Institut d'Astrophysique de Paris, 98bis Bd Arago, F-75014 Paris, France}
\and
{Istituto di Astrofisica Spaziale, CNR, Via Fosso del Cavaliere, I-00131 Roma, Italy}
\and
{Max-Planck Institut for Extraterrestrial Physics, D-85740 Garching, Germany}
\and
{Universit\`a degli Studi di Padova, Dipartimento di Astronomia, Vicolo dell'Osservatorio 5, I-35122 Padova, Italy}
\and
{Astrophysics Laboratory, Dept. of Applied Physics, Faculty of Engineering, Miyazaki University, 1-1 Gakuen-Kibandai-Nishi, Miyazaki 889-2192, Japan}
}
   
\offprints{M.Guainazzi, mguainaz@xmm.vilspa.esa.es}

\date{Received 10 November 1999 ; accepted 21 December 1999}

\maketitle

\markboth{M.Guainazzi et al.}{ A direct view of the AGN powering \src }

\begin{abstract}
We report the first direct X-ray evidence that an AGN is hidden in the 
center of \src. An ASCA observation of this target unveiled
a bright (0.5--10~keV luminosity $3.9 \times 10^{42}$~erg~s$^{-1}$)
and variable source, with minimum observed doubling/halving
time scale comprised in the
range 30--75~ks.
A model composed by a simple power-law, with photon index $\simeq$1.8
and an absorption edge, whose threshold energy is consistent with K-shell
photoionization of O{\sc vii}, provides an adequate fit of the spectrum.
This suggests that we are
observing the emission from the nuclear region through
a warm absorber
of ${\rm N_H \simeq}$ a few $10^{21}$~cm$^{-2}$. If it has internal dust
with Galactic gas-to-dust ratio, it could explain the lack of broad
H$_{\beta}$ emission, even in the episodic presence of a  broad 
H$_{\alpha}$ emission
line. Optical spectra obtained over several years show indeed
variations in the strength of this broad  H$_{\alpha}$ component. 
A distribution of dusty, optically thick
matter on spatial scales a few hundreds parsec, which does {\it not}
intercept the line of sight towards the nucleus, is probably required to
account simultaneously for the relative [OIII] luminosity deficit 
in comparison to the X-rays. The high IR to X-ray luminosity ratio
is most likely due to intense star formation in the 
circumnuclear region. \src\ might thus 
exhibit simultaneously nuclear activity {\it and} remarkable star formation.

\end{abstract}


\section{Introduction}

The correlation between the ROSAT All Sky Survey (RASS)
and the IRAS Point Source
Catalog unveiled the existence of many galaxies with X-ray luminosity in
the range $10^{42}$--$10^{43}$~erg~s$^{-1}$, which had not been
previously classified as Active Galactic Nuclei (AGN)
(Boller et al. 1992). Subsequent
optical observations indicated that a large fraction of them were previously
unrecognized Seyfert galaxies (Moran et al. 1994, 1996; Dennefeld et al. in
preparation).  However, the
nature of a few of them is far from being understood in terms
of simple nuclear activity and deserves further investigations. \src\ 
(NGC~4619) is one of those, which we started to study in more details in
1992.  

\begin{figure}
\begin{center}
\epsfig{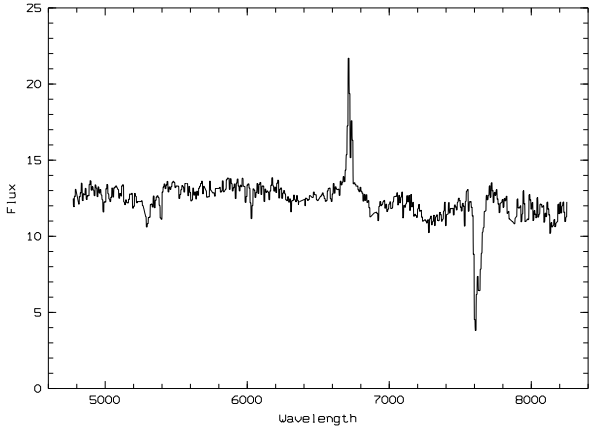}
\end{center}
\caption{March 1992 low resolution  spectrum of \src\ .}
\label{figmd1}
\end{figure}

It  is a nearby (${\rm z = 0.023}$)
barred spiral galaxy of morphological type SB(r)b, with a
luminous nucleus and relatively weak spiral arms. Our first optical spectrum
(Boller et al. 1993) 
was obtained at low resolution in March 1992 at the 1.93m telescope of 
Haute-Provence Observatory (OHP) and is displayed in Fig.~\ref{figmd1}.
It shows a red continuum, 
typical of IRAS galaxies, with many stellar absorption features and
conspicuous emission only in the red part of the spectrum (H$_{\alpha}$,
[N{\sc ii}]). H$_{\beta}$ is in absorption, and the [OIII] emission barely visible.
The existence of a strong NaD absorption line is an indication that
significant internal extinction is present.
No clear evidence is seen of a broad H$_{\alpha}$ component: a hint
of it may  be seen on the red side of the narrow emission line, but the
presence of the atmospheric B band does not allow a proper estimate of the
level of the continuum; this prompted 
further observations. \\
In the mean time, other spectra were published by Moran et al. (1994) and 
Mas-Hesse et al. (1996, hereinafter M96), showing a broad H$_{\alpha}$ 
component attributed to a Seyfert 1 nucleus. While their two
low-dispersion spectra are compatible with each other, their strong, broad 
H$_{\alpha}$ line is not seen on ours. A high-resolution spectrum taken 
by M96 only 25 days after our first spectrum displays a faint, H$_{\alpha}$ 
broad 
component, but apparently much fainter than in their low-dispersion spectrum  
taken about two years later. M96 noted also that 
the N{\sc ii}/H$_{\alpha}$ ratio  in \src\ was reminiscent of
a LINER (Filippenko 1993 and references therein), but that,
apart of it, the spectrum was remarkably featureless. 
From the lack of any detectable broad
component of the H$_{\beta}$, they estimated an amount of
extinction ${\rm E(B-V) \simgt 0.4}$, corresponding to
${\rm N_H \simgt 2 \times 10^{21}}$~cm$^{-2}$
(Lequeux et al. 1981; Prehdel \& Schmitt 1995); such a value is not
inconsistent with the reddening  estimated from the narrow components. 
The UV spectra obtained by M96 show also broad Ly$_{\alpha}$ and 
Mg{\sc ii} emission, as well as some absorption lines. 
The ${\rm E(B-V)}$ inferred from the UV measurements is $\approx$0.20.

The optical to X-ray Spectral Energy Distribution (SED) of \src\ has been
studied in detail by M96. They pointed out that
a scenario consisting of intense starburst
can well explain the infrared, optical and UV emissions and the
optical emission lines, but underestimates the soft X-rays. On
the other hand, if UV and X-rays are dominated by a non-stellar
contribution, the FIR emission is then highly in excess over the one typically
observed in Seyfert galaxies. The presence of both sources, with
comparable amount of energy output, seemed the most viable solution
for this object. Secular optical
variability in itself is not discriminating between the two
mechanisms if timescales are not determined, as Terlevich et al. (1992) 
have argued that massive star formation could also account both for
variability  and broad lines.

\src\ is bright in soft X-rays [${\rm \log(L_{0.1-2.4 keV}) = 42.80}$],
with a photon spectral index ${\rm \Gamma \simeq 1.90}$
(Moran et al. 1996).
Studies in the X-rays are in principle of the uppermost importance to
address properly the true origin of the output power,
because they can pierce throughout the
innermost regions of the galaxy, and unveil the high-energy processes
that occur in the immediate proximity of the nucleus.
An ASCA observational program was started to search for evidence of
nuclear activity in a sub-set of the Boller et al. (1992)
soft X-ray luminous galaxies sample.
In this paper, we report the results
on \src\ (the only target for which
time was allocated), which provided the first hard X-ray measure for
this object. The results are presented together with a re-analysis of a
pointed archival ROSAT/PSPC observation and with new optical data obtained
in the mean time. We will assume
${\rm H_0 = 50}$~km~s$^{-1}$~Mpc$^{-1}$.
Energies are quoted in the source rest frame and
errors are at 90\% confidence level for one interesting parameter
(${\rm \Delta \chi^2 = 2.71}$).

\section{X-ray Data Reduction and analysis}

\src\ was observed by ROSAT/PSPC and ASCA on 1993, 16--17 December, and
1997, 4--5 June, respectively. The data have been retrieved from
the HEASARC archive, as cleaned event lists. 
The ROSAT PSPC is a proportional counter sensitive in the 0.1--2~keV band,
with moderate energy resolution (${\rm \Delta E/E \simeq 50\%}$ at 1~keV)
and a spatial resolution of about 20$\arcsec$ for on-axis sources.
The ASCA
scientific payload comprises two
Solid Imaging Spectrometers (SIS0 and SIS1, sensitive in the 0.6--9~keV band,
Gendreau 1995)
and two Gas Imaging Spectrometers (GIS2 and GIS3, sensitive in the 0.7--10~keV
band, Ohashi et al. 1996). SIS grades 0,1,2 and 4 and 1-CCD mode were used.

\begin{figure}
\begin{center}
\epsfig{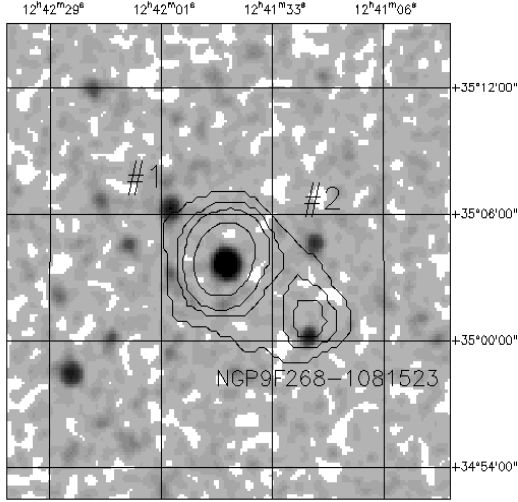}
\end{center}
\caption{PSPC field of view of \src\ (the central source)
superimposed to the GIS2 intensity contours of the ASCA observation}
\label{fig3}
\end{figure}

The best ROSAT/PSPC centroid position
($\alpha_{2000.0}$=$12^{h}41^{m}44^{s}.8$ and $\delta_{2000.0}$=$35^{o}$03$\arcmin$46$\arcsec$)
is within 22$\arcsec$ from the optical position (Santagata et al. 1987)
and 40$\arcsec$ from the ASCA centroid. The latter quantity is
comparable with the typical ASCA positional uncertainties (Gotthelf
\& Ishibashi 1997).
Several serendipitous sources are present in the PSPC field of view (see
Fig.~\ref{fig3}).
Two of them could
in principle contaminate
the ASCA source spots. Their properties are reported in Table~\ref{tab3}.
\begin{table*}
\begin{footnotesize}
\begin{center}
\begin{tabular}{lcccccc} \hline \hline
Source & ${\rm \alpha_{2000}}$ & ${\rm \delta_{2000}}$ & ${\rm N_H}$ & ${\Gamma}$ & 0.5--2~keV flux & 2--10~keV flux$^a$ \\
& & & ($10^{20}$~cm$^{-2}$) & & ($10^{-14}$~erg~cm$^{-2}$~s$^{-1}$) & ($10^{-14}$~erg~cm$^{-2}$~s$^{-1}$) \\ 
\#1 & $12^{h}41^{m}58^{s}.2$ & $35^{o}$06$\arcmin$24$\arcsec$ & $2.0 \pm^{1.6}_{1.8}$ & $2.7 \pm 0.5$ & 5 & 6 \\
\#2 & $12^{h}41^{m}58^{s}.8$ & $35^{o}$03$\arcmin$21$\arcsec$ & $< 1.2$ & $2.2\pm^{0.8}_{0.5}$ & 1.7 & 1.5 \\ \hline \hline
\end{tabular}
\end{center}

\noindent
\hspace{0.75cm}$^a$extrapolated from the best-fit PSPC model

\caption{Properties of the serendipitous sources \#1 and \#2 in the PSPC field. Spectra parameters refer to a power-law plus photoelectric absorption model of the PSPC spectrum.}
\label{tab3}
\end{footnotesize}
\end{table*}
We expect that
any contamination
to the \src\ ASCA flux is
lower than 10\% and 5\% in the 0.5--2 and 2--10~keV bands, respectively,
well within the statistical uncertainties of the ASCA measurement.
PSPC scientific products have been extracted from a circular region of
about 2$\arcmin$ around the apparent centroid of the source. Background
spectra have been extracted in an annulus of radii 3$\arcmin$20$\arcsec$ and
5$\arcmin$20$\arcsec$, after removing circular areas of 2$\arcmin$ radius around any
serendipitous source included in the annulus.

In the SIS0 or SIS1 images no source is detected other than \src,
the 3$\sigma$ upper limit at the position of source \#1 in PSPC field being
$8 \times 10^{-3}$~s$^{-1}$.
The closest serendipitous source in the GIS field of view
($\alpha_{2000.0}$=$12^{h}41^{m}44^{s}.8$ and
$\delta_{2000.0}$=$35^{\circ}$03$\arcmin$46$\arcsec$;
distance 5.8$\arcmin$) is probably associated with the foreground galaxy
NGP9 F268-1081523.
ASCA source scientific products have been then extracted from circles of
radii 4$\arcmin$, 3$\arcmin$.25 and 3$\arcmin$.75, 
in the SIS0, SIS1 and GIS, respectively.
Background scientific products have
been extracted from regions of the field of view free from contaminating
sources. None of the presented results changes significantly if the
background is extracted from blank sky fields.
Total exposure times are about 33, 32 and 13~ks for the SIS, GIS and PSPC,
respectively. Full band count rates were $(3.79 \pm 0.12)$, $(2.77 \pm 0.11)$,
$(1.89 \pm 0.09)$, $(2.38 \pm 0.10)$ and $(6.4 \pm 0.3) \times 10^{-2}$s$^{-1}$,
for the SIS0, SIS1, GIS2, GIS3 and PSPC, respectively.

\subsection{The X-ray light curve}

\begin{figure}
\begin{center}
\epsfig{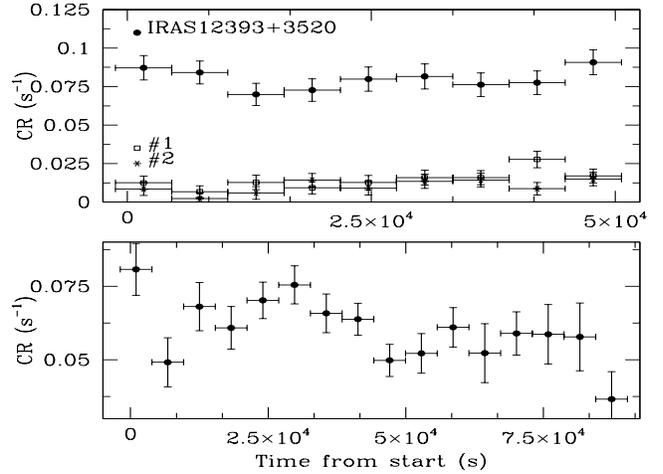}
\end{center}
\caption{0.1--2~keV background-subtracted
(ROSAT/PSPC, {\it upper panel}) and 0.5--4~keV
(ASCA/SIS, {\it lower panel}) light curves. Binning time is 5760~s.
Only bins with an exposure fraction higher than 15\% are shown. If a
linear fit is performed on the ASCA light curve in
the 5--35 and 35--50~ks intervals, one obtains a rate of flux change
of ${\rm 40 \pm 30 \%}$ and ${\rm 40 \pm 20 \%}$, respectively}
\label{fig1}
\end{figure}

In Fig.~\ref{fig1} the light curves in the 0.1--2~keV (PSPC) and
0.5--4~keV bands (SIS) are shown.
The latter shows
a gentle rise and decay before about 50~ks from the start of the observation.
A similar trend is observed in the GIS light curves.
If a fit with a constant line is performed on the light curve,
${\rm \chi^2 |_{ASCA} = 30/15}$ degrees of freedom
(dof). From the flux change rate of the light curve in the
5--35 and 35--50~ks intervals,
we estimate a minimum doubling/halving time in the range 30--75~ks.
Variability on lower timescales cannot be ruled out, but the
limited available statistics prevented us to reach a firm conclusion on its
existence.
This variability is not associated with any of the serendipitous sources
detected in the PSPC and laying at the border of the ASCA extraction regions.
We have extracted SIS0 images in the time intervals between
10-40 and 50-80~ks after the beginning of the ASCA observation.
The source spot does not show sign of
elongation and the centroid best-fit positions agree within 6$\arcsec$.
Although an eye inspection of the PSPC
light curve might suggest that a similar variability was
observed by ROSAT as well, the evidence is not statistically that
clear (${\rm \chi^2 |_{ROSAT} = 18/21}$~dof).
We have  searched for spectral dynamics associated with the ASCA
flux changes, without success. We will therefore focus in the next section 
on the time averaged spectrum of \src.

\subsection{Spectral analysis}

All spectra have been rebinned in order to have at least 20 counts per
energy channel, in order to ensure the applicability of the $\chi^2$
statistics. Proper matrices for the date of the observations have been
retrieved from the HEASARC archive or
built with the software available in the {\sc Ftools 4.0} version.
First, we fitted the PSPC and ASCA spectra in the overlapping
0.8--2~keV band
with a simple photoelectric absorbed power--law model,
leaving all parameters free (except the ${\rm N_H}$, which has been
tight to be the same for all instruments), to check if
significant spectral variability arose between the epochs of
the two observations. The spectral indices
turn out to be well consistent within the statistical
uncertainties (${\rm \Gamma^{0.8-2 keV}_{ROSAT} =
1.7 \pm^{1.0}_{0.8}}$; ${\rm \Gamma^{0.8-2 keV}_{ASCA} =
1.5 \pm^{0.7}_{0.4}}$), despite of an increase in the
flux of $\simeq 130 \%$. In the following, we will therefore fit
the spectra of all detectors simultaneously, only allowing
a relative normalization factor as a free parameter among all
the instruments.

\begin{figure}
\begin{center}
\epsfig{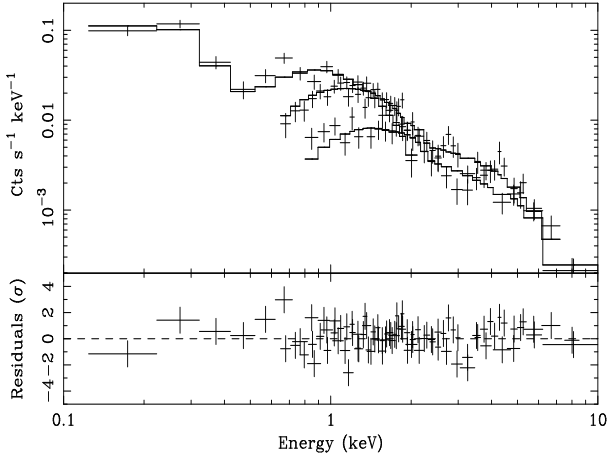}
\end{center}
\caption{PSPC, SIS0 and GIS2 spectra ({\it upper panel}) and
residuals in units of standard deviations ({\it lower panel})
when a photoelectric absorbed power-law model is applied to
the data of all detector simultaneously. Each data points has a
signal-to-noise ratio $>3$}
\label{fig2}
\end{figure}

In Fig.~\ref{fig2}, the results of a spectral fit with a simple
power-law with photoelectric absorption are shown.
The fit is
rather good (${\rm \chi^2 = 188/177}$~dof), with ${\rm \Gamma =
1.55 \pm^{0.08}_{0.05}}$ and
${\rm N_H = (0.8 \pm^{0.3}_{0.2}) \times 10^{20}}$~cm$^{-2}$.
Further spectral complexity is not strongly required by the data.
The only feature, whose addition is required at $\simeq$99\% level,
is an absorption edge (${\rm \Delta \chi^2 = 9}$ for two more
parameters), with threshold energy ${\rm E_{th} = 0.71 \pm^{0.05}_{0.06}}$,
formally consistent with the K-shell photoionization energy of O{\sc vii}.
The edge is detected with comparable significance in both the ROSAT
and ASCA spectra separately.
Similar features have been commonly observed in the spectra of Seyfert~1
galaxies (Reynolds 1997; George et al. 1998).
The addition of this feature makes the spectrum steeper and therefore
consistent with that typically observed in Seyfert 1 galaxies
(Nandra \& Pounds 1994; Nandra et al. 1997b; see Table~\ref{tab1}).
\begin{table*}
\begin{footnotesize}
\begin{center}
\begin{tabular}{lccccccc} \hline \hline
Model & ${\rm N_H}$ & ${\rm \Gamma}$ & ${\rm R}$ & ${\rm E_{th}}$ & ${\rm \tau}$ & ${\rm \chi^2/}$~dof \\
& (${\rm 10^{20}}$~cm$^{-2}$) & & & (keV) & &  \\
{\verb!wa*ed*(px+ga)!} & $1.6 \pm^{0.5}_{0.4}$ & $1.78 \pm^{0.12}_{0.11}$ & 0$^{\dag}$ & $0.71 \pm^{0.05}_{0.06}$ & $0.7 \pm^{0.4}_{0.3}$ & 172.4/175 \\
{\verb!wa*ed*(px+ga)!}& $1.7 \pm^{0.5}_{0.4}$ & $1.82 \pm^{0.12}_{0.11}$ & $1^{\dag}$ & $0.71 \pm^{0.04}_{0.06}$ & $0.8 \pm 0.4$ & 171.0/175 \\
{\verb!wa*ed*(px+ga)!}& $2.3 \pm^{1.0}_{0.7}$ & $2.0 \pm^{0.3}_{0.2}$ & $7.0 \pm^{9.0}_{5.7}$ &  $0.69 \pm^{0.05}_{0.04}$ & $1.0 \pm^{0.5}_{0.4}$ & 167.7/174 \\ \hline
Model & ${\rm N_H}$ & ${\rm kT_1}$ & ${\rm kT_2}$ & ${\rm Z_1 = Z_2}$ & ${\rm \chi^2/}$~dof \\
& (${\rm 10^{20}}$~cm$^{-2}$) & (eV) & (keV) & \% &  \\
{\verb!wa*(mk+mk+ga)!}& $3.1 \pm 1.8$ & $110 \pm 40$ & $8.1 \pm^{2.5}_{1.6}$ & $0.6 \pm^{0.5}_{0.4}$ &  181.4/175 \\ \hline \hline
\end{tabular}
\end{center}

\noindent
\hspace{2.5cm}$^{\dag}$fixed

\caption{Best-fit parameters and results. {\it wa}~=~photoelectric absorption;
{\it ed}~=~photoelectric absorption edge; {\it px}~=~Compton reflection;
{\it mk}~=~optically thin plasma; {\it ga}~=~Gaussian emission line}
\label{tab1}
\end{footnotesize}
\end{table*}
We have therefore performed a fit with a Seyfert-like model,
constituted by a Compton-reflected power-law (model {\verb!pexrav!} in
{\sc Xspec}, Magdziarz \& Zdziarski 1995), an absorption edge and
a fluorescent broad (i.e.: $\sigma = 0.43$~keV, Nandra et al. 1997b)
iron line from neutral iron (i.e: centroid energy held fixed to 6.4~keV).
The line is actually not required by the fit, but we have included it
because it is expected on theoretical grounds (George \& Fabian 1991;
Matt et al. 1992). The fit is comparably good as in the simple power-law
case.  The amount of reflection, parameterized through
the ratio between the reflected and the transmitted components, R\footnote{R
is equal to 1 when the reflection occurs in a plane-parallel infinite slab.
A higher value might imply either a higher solid angle subtended by the
disk to the source -
as {\it e.g.} in a disk ``warped'' geometry -,
a delay in the disk response to flux
changes of the primary continuum, or an anisotropy of the nuclear
emission, such that the disk sees more flux than emitted along our
cone-of-sight},
is $>$1.3, which is not inconsistent with
the upper limit on the EW of the iron line (600~eV),
if standard cosmic abundances are assumed
(Matt et al. 1992).
If we add to this model a 1~keV thermal emission from an optically
thin plasma, its 0.1--10~keV luminosity is constrained to be
lower than ${\rm 3.8 \times 10^{41}}$~erg~s$^{-1}$ at the 90\% confidence
level.
The cold absorbing column density is
broadly consistent with
the Galactic contribution along the \src\ line of sight (${\rm N_{H,Gal} =
1.4 \times 10^{20}}$~cm$^{-2}$, Dickey \& Lockman 1990).
The fluxes in the 0.5--2, 0.5--4.5 and 2--10~keV energy bands
are ${\rm 5.2 \times 10^{-13}}$, 1.00 and
${\rm 1.12 \times 10^{-12}}$~erg~cm$^{-2}$~s$^{-1}$, implying
rest frame unabsorbed luminosities of 1.29, 2.4
and ${\rm 2.6 \times 10^{42}}$~erg~s$^{-1}$, respectively.

Alternatively, a good fit can be formally
obtained also with a double-temperature
optically thin plasma (code {\verb!mekal!} in {\sc
Xspec}), with temperatures $\simeq 100$~eV and $\simeq 8$~keV
and abundances consistent with solar. In this scenario, the contribution
of a $\Gamma = 1.9$ (1.0) power-law to the 0.5-4~keV flux is
lower than 7\% (6\%).

\section{Additional optical data}

\begin{figure}
\begin{center}
\epsfig{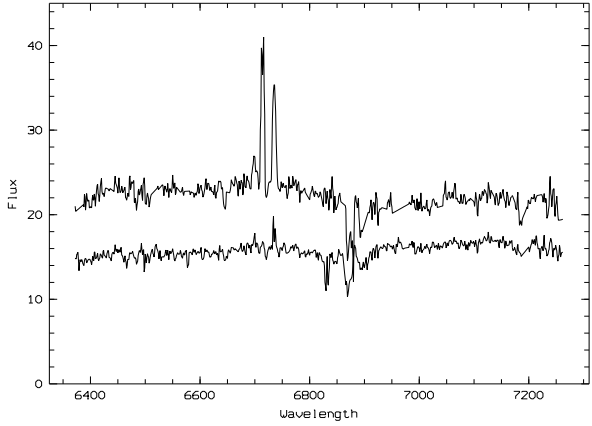}
\end{center}
\caption{January 1996 \src\ high dispersion spectra, with extraction integrated
 along 40$\arcsec$ ({\it upper}) and 10$\arcsec$ ({\it lower}) slit lengths.}
\label{figmd2}
\end{figure}

An intermediate resolution spectrum (1.8 \AA per pixel) of the H$_{\alpha}$ 
region of NGC 4619 was obtained in January 1996 at OHP. 
The calibrated one dimensional spectrum has been extracted in two ways,   
shown in Fig.~\ref{figmd2}.
The first one (upper spectrum in Fig.~\ref{figmd2}) integrates 
over all the length were 
H$_{\alpha}$ is detected (about 40"): the emission lines are  clearly
detected, with a [N{\sc ii}]/H$_{\alpha}$ ratio stronger than in standard 
H{\sc ii} regions. This ratio was the basis for the LINER claim in M96. 
Unfortunately, the
[S{\sc ii}] lines, which could provide an additional diagnostic, fall into an
atmospheric absorption line. As any clear broad component, revealing an AGN, 
could be hidden in a much larger star forming region, a second extraction 
was performed over a smaller extension (about 10"), corresponding to the
strong continuum. This is the lower spectrum of Fig.\ref{figmd2}, where 
only weak [N{\sc ii}] emission lines
are seen, and almost no H$_{\alpha}$. As the latter may however be hidden in 
the corresponding stellar absorption, it is difficult to use 
the  [N{\sc ii}]/H$_{\alpha}$ line ratio alone to claim for
a diagnostic of Seyfert or LINER. In both spectra  a broad H$_{\alpha}$ 
component may  be marginally seen but, if present, is certainly not as strong 
as seen in the high dispersion spectrum of M96 (and even less so than in 
their low dispersion one).

\begin{figure}
\begin{center}
\epsfig{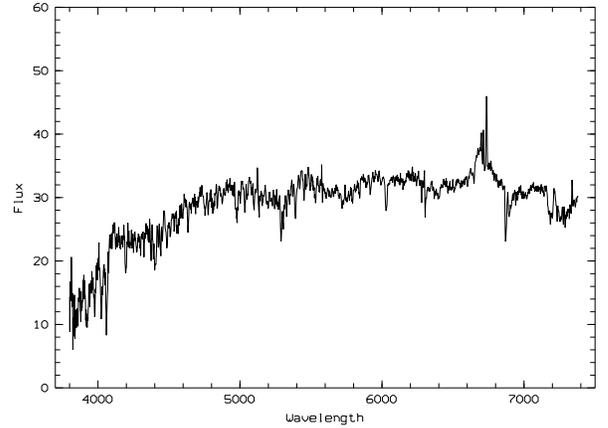}
\end{center}
\caption{June 1999 low dispersion  spectrum of \src. }
\label{figmd3}
\end{figure}

We finally obtained a new, low dispersion spectrum, in June 1999 at OHP, and
the result is shown in Fig.~\ref{figmd3}.
This spectrum, extracted over a similar 10" 
width as the lower one in Fig.~\ref{figmd2},
shows a clear, broad, asymmetric wing to 
H$_{\alpha}$, with a strong component on the blue side (where nothing was
suspected in our 1992 spectrum) very similar to the 
one seen in the low dispersion spectra of
M96 and Moran et al. (1994). While a quantitative comparison between the
various spectra is difficult, as slit orientations and extraction lengths
are different (and not always well documented), we have here reasonable 
evidence for variations of a broad-line component in a spectrum otherwise 
dominated by intermediate-type stars. The broad H$_{\alpha}$ component is 
absent or weak in the first available spectra (ours in March 1992, the 
high-resolution one of April 1992 in M96), is strong in the low dispersion 
spectra of Moran et al. (1994) and of M96 (July 1994), dimmed again in our 
data of January 1996 and strong again in our last spectrum in June 1999. The  
variations seen are between a Seyfert 1.9 spectrum and 
a spectrum dominated by  stellar features. Some indications exist for a LINER
identification: 
we see a strong 
[N{\sc ii}]/H$_{\alpha}$ ratio, as M96 did, and our 1999 spectrum shows also 
an [O{\sc ii}]/[O{\sc iii}] ratio much greater than one. We have however no
detection of the [O{\sc i}] line with a strength which could confirm the LINER
diagnostic (Heckman, 1980). We note that up to now, only one case is
known where broad lines developed in a LINER (NGC 1097; Storchi-Bergmann et
al. 1993). The substantial reddening derived in 
our last spectrum [${\rm E(B-V)\simeq 0.8-0.9}$, consistent between the Balmer
decrement in the narrow component and  the NaD absorption], 
 is certainly an element of importance in the following discussion.

\section{Discussion}

\subsection{The hard X-ray properties of \src }

X-rays provide a strong evidence in favor of the existence of an AGN
in \src. We have detected a significant variability of the
X-ray light curve. The minimum observed associated doubling/halving
time is comprised in the range 30--75~ks.
If the variability is intrinsic to the radiation emitted
in the nuclear region of an AGN, usual light crossing
arguments constrain the mass of the nuclear black hole
to be ${\rm < 10^9 \eta^{-1}_{5} M_{\odot}}$,
where ${\rm \eta_5 \equiv r/(5 R_S)}$ is the typical size of
the region emitting the bulk of the primary non-thermal
continuum in unit of five Schwarzschild radii.
The joint ROSAT/ASCA 0.1--10~keV
spectrum has also a shape which closely resembles the one
typically observed in radio-quiet AGN.
Our data suggest that the solid angle subtended by the accretion disk
is higher than due to a plane-parallel infinite slab.
The upper limit on the EW of a fluorescent neutral iron line (600~eV) is
not inconsistent with this scenario if solar abundances are assumed.
However, much better quality
of the data is required to confirm this hint.

The data require at 99\% level of confidence the presence of an
absorption edge, whose threshold energy is
consistent with the K-shell photoionization energy of O{\sc vii}.
In at least 50\% of Seyfert 1s
the nuclear radiation is absorbed by substantially ionized matter
(Reynolds 1997; George et al. 1998). In \src\
the properties of the absorbing matter are
statistically poorly constrained and are somewhat dependent on the
assumed continuum model. We conservatively derive a column
density in the range 2.5--$7.5 \times
10^{21}$~cm$^{-2}$ [or ${\rm E(B-V) \simeq 0.5}$--1.5 under standard
assumptions],
if the physical conditions
required to sustain an oxygen state dominated by He-like are assumed.
If the warm absorber contains
dust in the Galactic dust-to-gas ratio,  such a medium
may account for the observed
optical reddening and explain the missing H$_{\beta}$ broad line, even
in the presence of a strong broad H$_{\alpha}$
(which is actually detected only in some of the \src\
optical spectra, cf. Sect.~3). ``Dusty'' warm
absorbers have already been invoked to explain the discrepancy between the
amount of X-ray cold absorption and the optical reddening in a few Seyfert
galaxies (Reynolds et al. 1997; Komossa \& Fink 1998 and
references therein), although 
Siebert et al. (1999) have recently shown that such a "simple" representation 
was not able to reproduce all the observed optical and X-rays data in the
case of a well studied example,  IRAS13349+2438. 

The observed spectra can be formally accounted by a two-temperature
optically thin plasma, in analogy with recent evidence on the X-ray
emission of starburst galaxies (Della Ceca et al.
1998; Cappi et al. 1999).
In this scenario, however, the observed short-term variability
cannot be explained, unless it represents the diluted appearance of
a much more variable underlying unresolved source.
The lack of spectral variability during the ASCA observation points
against the emerging of a further source in the spectrum in
correspondence to the flux variation.
The contribution of a power-law source in the 0.5-4~keV band is
constrained to be lower than $1.8 \times 10^{41}$~erg~s$^{-1}$. To
reproduce the observed flux change, it should vary
in $10^4$~seconds by an amount $\simeq \frac{12}{L_{41}}$,
where ${\rm L_{41}}$ is the X-ray luminosity in units of
${\rm 10^{41}}$~erg~s$^{-1}$. This rules out the
possible serendipitous detection
of local X-ray binaries or superluminal sources; the latter ones
exhibit indeed variability in flux by an amount
up to two orders of magnitude, but have normally luminosities well
within $10^{39}$~erg~s$^{-1}$. On the other hand, X-ray variability
on such timescales is rather common in Seyfert galaxies
(Nandra et al. 1997a and references therein). The timescales are too fast
to be explained by mechanisms which do not invoke accretion on a supermassive
black hole ({\it e.g.} the starburst model of Terlevich et al. 1992).

\subsection{Comparison between X-ray and other wavelengths}

\begin{figure}
\begin{center}
\epsfig{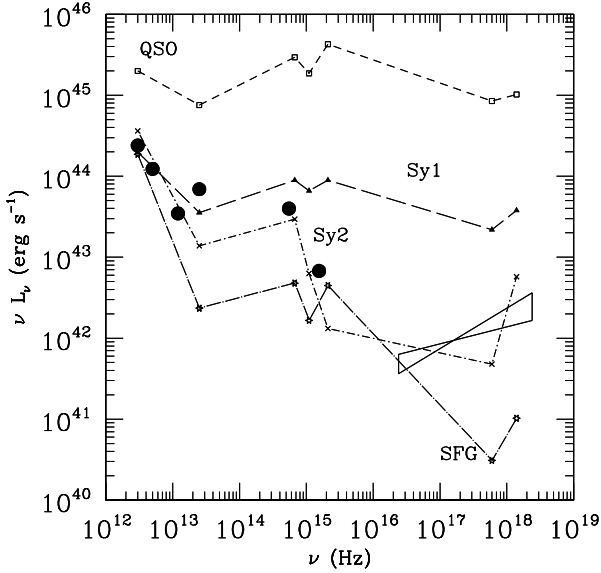}
\end{center}
\caption{\src\ SED ({\it filled circles and bowtie}) compared with the
average ``tracks'' observed in quasars (QSO, {\it empty squares}), Seyfert~1s
(Sy1, {\it filled triangles}), Seyfert~2s (Sy2, {\it crosses}) and
star forming galaxies (SFG, {\it empty stars}; Mass-Hesse et al. 1995).
IR to UV \src\ photometry points are from M96.}
\label{fig5}
\end{figure}

\begin{figure}
\begin{center}
\epsfig{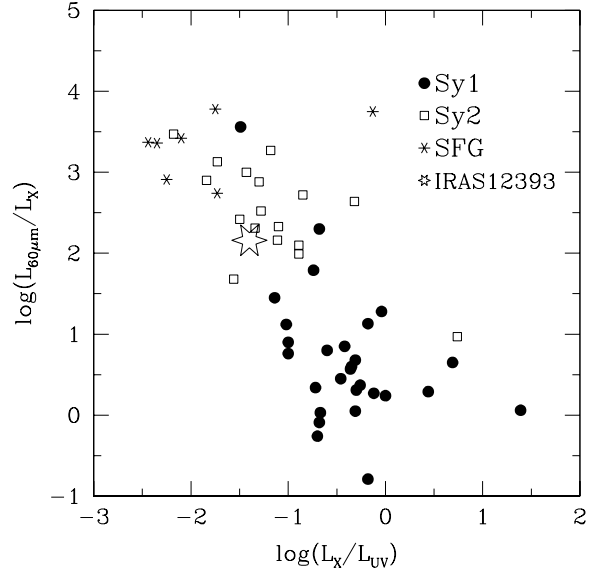}
\end{center}
\caption{IR (60 $\mu$m), UV (2700 ${\rm \AA}$) and X-ray (0.5--4.5~keV)
color-color diagram for Seyfert~1s ({\it filled
circles}), Seyfert~1.5--2 ({\it empty squares}) and SFG ({\it stars}). The
location of \src\ is indicated by the big empty star. Data are from Mass-Hesse et al. 1995}
\label{fig4}
\end{figure}

An energetic argument points against the idea that hard X-rays are dominated
by a nuclear starburst.
No known starburst galaxy is so X-ray luminous (Ptak et al. 1999).
The soft-X vs. 60$\mu$m and soft-X vs. UV luminosity ratios are
$1.9 \times 10^{-2}$ and $1.2 \times 10^{-2}$,
more than one order of magnitude higher than typical
values observed in star forming galaxies (SFG, Mass-Hesse et al. 1995).
The observed X-ray luminosity $\sim 4 \times 10^{42}$~erg~s$^{-1}$ is,
on the other hand, 
not uncommon among Seyfert galaxies. In Figure~\ref{fig5}
the \src\ SED is shown, superimposed to the average ``tracks'' of the
Mass-Hesse et al. (1995) sample for Seyfert~1s, Seyfert~2s, QSO and SFG.
Caution must be employed when evaluating the SED in \src, because it
refers to non-simultaneous measurements. Optical spectroscopy suggests
that the nuclear activity (as seen by the BLR), and therefore also the
energy budget, might be variable with time. Given these caveats, one notes
that \src\ follows apparently better  
a Seyfert~2 SED (or eventually a SFG one), being strongly under-luminous in
UV and X-rays in comparison to the IR, if the normalization is done in the 
far-IR.
These pieces of information are summarized in the IR, UV and X-ray
color-color plot of Figure~\ref{fig4}, where \src\ is seen  to lie in the 
region of Seyfert galaxies, and apparently type 2 rather than 1. 
The X vs. [OIII] luminosity ratio is $3.7 \times 10^2$,
an order of magnitude greater than observed in Seyfert~1s 
(cf. Fig.~6 in Maiolino et
al. 1998) and would also point towards a Seyfert 2 identification. 

Several pieces of evidence point, however, against an identification of  
\src\ as a Seyfert~2. First, 
there is no evidence of the high excitation typical of Sey2's: 
the [OIII] over H$_{\beta}$ line ratio is difficult to measure accurately, as 
[OIII] is faint and  H$_{\beta}$ is hidden in the corresponding stellar
absorption, but is not larger than three. In addition, the [OII]/[OIII] ratio 
is much larger than one, and it is the combination of these  line
ratios, with the [NII]/H$_{\alpha}$ one, which lead to the LINER claim for 
this object. This is not compatible with a Seyfert 2 identification.
 Moreover,  a high excitation 
is not seen either when the broad H$_{\alpha}$ component is absent, so that 
the variations cannot be described as a transition between a Seyfert 1 and a 
Seyfert 2 phase, as seen in some other cases (e.g. Mkn~993, Tran et al.
1992; Mkn~6, Khachikian \& Weedman 1971; Mkn~1018,
Cohen et al. 1986). Second, the X-ray intrinsic neutral absorption
column density is negligible.
No model where the nuclear power-law is strongly absorbed
yields comparably good fits to the ROSAT/ASCA spectrum as the
simple models of Table~\ref{tab1}, even if one allows the covering
fraction of the absorbing matter to be lower than 1.
One might suppose that the X-ray spectrum is totally dominated by scattered
radiation from an otherwise invisible nucleus, as observed in several
``reflection-dominated'' Seyfert~2s (Turner et al. 1997; Matt et al. 1997;
Guainazzi et al. 1999).
The observed variability, however, introduces severe constraints on 
 this possibility. Assuming a typical variability timescale of
${\rm \Delta t \approx 4 \times 10^4}$~s,
light crossing arguments imply a lower limit on the
electron numerical density ${\rm n_e \simgt (c \Delta t \sigma_{sc})^{-1}
\sim 1.2 \times 10^9}$~cm$^{-3}$. If the absorption occurs in the same
medium, an optical depth to scattering of the order of one implies
an optical path
$\sim N_H/n_e \simlt 4 \times 10^{12}$~cm. Such a low value
would imply that the scattering/absorption occurs in the proximity of the
galactic nucleus, hardly compatible with the idea that the matter hiding the
nuclear region is located at distances $\sim$1--100~pc, in the shape of a,
more or less homogeneous, azimuthally symmetric structure (Antonucci \&
Miller 1985; Maiolino \& Rielke 1995; Greenhill et al. 1996).
Finally, the appearance sometimes of a broad H$_{\alpha}$ 
component is a definite sign in favor of a Seyfert 1 classification. 

One clue to explain the observed SED is
to assume that the AGN is a weak Seyfert~1, which is  over-luminous
in IR in comparison to the objects of the same class of comparable
X-ray luminosity. This
interpretation can be naturally linked to the puzzling absorption pattern
emerging from the optical/UV
spectroscopy. Very broad
Ly$_{\alpha}$ and H$_{\alpha}$ lines have been observed in (not simultaneous)
observations of the core of \src\ (M96).
All these evidences suggest that the geometrical distribution of the
circumnuclear neutral absorbing matter
is far more complex than the simple equatorially symmetric pattern, which
is assumed in the 0-th order Seyfert unification theories (Antonucci \&
Miller 1985; Antonucci 1993). The Narrow Line Region (NLR) could be absorbed
by dusty, optically thick matter distributed on spatial scale of a few
hundreds of parsec, which does not (not always?) intercept the line of sight 
towards the nuclear environment.
This dust might simultaneously be responsible
for the IR emission through reprocessing of the incoming nuclear
radiation during particularly active phases (or when the
line of sight between the nucleus and the dust is unobscured).
In the same framework, the
under-luminosity in ${\rm L_{[OIII]}}$ and the relative over-luminosity
in IR in comparison to the directly observed X-ray of nuclear
origin can be simultaneously explained. A similar scenario
had been originally suggested
by Maiolino \& Rieke (1995), who proposed that ``intermediate'' Seyfert
galaxies are seen through a 100~pc-scale torus coplanar with the plane
of the galaxies.
This idea has received a further support after the results
of a recent HST slew survey, which lead to the discovery
that the distribution of matter in the nuclear environment of
radio-quiet, nearby AGN is indeed highly patchy, with dusty
lanes protruding from several kilo-parsecs to a few hundreds of parsecs
towards the
center (Malkan et al. 1998).
Recently, Maiolino et al. (1999) pointed out that barred galaxies
(like \src) tend to exhibit the highest values of IR to X-ray
luminosity ratio (see, however, a different point of view in
Regan \& Mulchaey (1999). This might suggest that stellar bars are very
efficient in driving gas to the circumnuclear region and therefore
to provide high amount of warm (AGN-heated) dust in the nuclear
environment, which would reprocess the nuclear high-energy continuum.

The bulk of the IR radiation could  alternatively be produced by an
intense star formation episode, occurring on spatial scales of the
order or larger than the NLR. The SED in the IR alone is typical of 
starburst galaxies and does not satisfy the various criteria defined to 
select AGN in IRAS data (de Grijp et al. 1985; D\'esert \& Dennefeld 
1988). Also, the standard IR/radio parameter q, discussed by Condon at al.
(1995), has 
a value of 2.85 for \src, showing that the AGN, if present, is not dominating 
the IR emission and/or that the object is radio-weak. Finally, the
observed infrared luminosity
(${\rm L_{IR} \sim 5 \times 10^{44}}$~erg~s$^{-1}$; M96) is almost two orders
of magnitude higher than expected from the AGN 1--10~keV luminosity
(${\rm L_X \sim 3 \times 10^{42}}$~erg~s$^{-1}$), if the \src\ AGN SED has
a ${\rm L_{IR}/L_X}$ ratio typical for quasars with
${\rm L_X < 10^{45}}$~erg~s$^{-1}$ ($\simeq$4.5; Elvis et al. 1994).

While both explanations for the IR emission are plausible, the energetic
argument just  
discussed  favors the hypothesis that the 
IR emission is dominated by star-formation. However, even in this case,  an 
additional (nuclear) source of high-energy radiation is required to explain
the hard X-rays. As shown in 
Fig.~\ref{fig4}, \src\
exhibits indeed a much higher X-ray to IR or UV luminosity flux
than typically observed in SFG galaxies.  It is likely to be
provided by a separate component, which is straightforward to identify,
in the light of the ASCA/ROSAT results, with an active nucleus. On the
other hand, a spatially and physically structured and time varying absorber 
is also required to yield the observed
different absorptions towards the nucleus, of the
BLR and of the NLR lines. The dusty environment of a star forming region
could provide only the last. \src\ is therefore best described by a central, 
weak, AGN, with a BLR partly obscured by a structured absorber, and a well 
absorbed NLR, mixed with a region of intense star formation of perhaps larger 
extension. 

\begin{acknowledgements}
  
MG acknowledges the receipt of an ESA Research Fellowship. The authors
acknowledge discussions with F.Bocchino. This research
has made use of the NASA/IPAC Extragalactic Database, which is operated
by the Jet Propulsion Laboratory under contract with NASA, and of data
obtained through the High Energy Astrophysics Science Archive Research
Center Online Service, provided by the NASA/Goddard Space Flight Center.

\end{acknowledgements}

\end{document}